# HIGH POWER SUPERCONDUCTING ELECTRON LINEAR ACCELERATORS FOR INDUSTRIAL APPLICATION

Jayakar Thangaraj*
Fermi National Accelerator Laboratory
P O Box 500, Batavia IL
jtobin@fnal.gov

**ABSTRACT**

Fermilab is executing a technology development program to develop a compact yet powerful electron accelerator. We are leveraging R&D breakthroughs in SRF cavities, cost-effective radio-frequency sources, modern cryo-coolers, and high average current electron guns. We show that a single accelerator module can deliver average beam power as high as 1 MW through detailed thermal, RF, and particle simulations.

## 1. INTRODUCTION

There is a need for robust, simple, high-average power, electron beam sources in many applications. Such electron beam sources could also replace Cobalt-60 for industrial x-ray inspection, medical sterilization, and food irradiation. Unlike radioactive sources, the radiation from accelerators has a switch that turns off the associated radiation. Therefore, there is the desire for turn-key systems designed and engineered to be sufficiently robust and reliable for commercial applications. Transformational technological advances in SRF and peripheral equipment pave the way for creating a viable, compact, powerful, high-power, high-energy electron beam source. Niobium cavities coated with $Nb_3Sn$ can be operated with gradients of > 10 Megavolts/m with a quality factor at a temperature of 4.2 K of >1 x $10^{10}$. Recent results and details are in references [1-3]. A cavity can be operated with Continuous Wave (CW) RF power while cooled by a single cryocooler. The very high Q0 using $Nb_3Sn$ coated 650 MHz cavities mean that the dynamic heat loads per cavity are sufficiently small that for the first time, SRF cavities can be cooled via conduction without being immersed in a liquid helium bath. [1]. Moreover, cryocoolers are widely used to supply 4K refrigeration to hospital MRI magnets and are rapidly becoming available at ever-larger capacities and lower prices. Cryocoolers eliminate the need for large cryogenic refrigerators, pressure vessels, helium gas/liquid inventory management systems, and expert operators. The overall result is dramatically simplifying operation, lower cost, and significantly improved reliability.

Using all these advancements, Fermilab is building complete turn-key, industrial, SRF accelerator systems that are highly reliable and operable by non-experts. While initial developments use the external injection and thermionic gun, other developments include field emission electron cathodes. These cathodes allow direct insertion into an SRF cavity, creating a very short and compact accelerator [4]. Also, using magnetrons to drive a narrow-band load like an SRF cavity can dramatically lower the capital cost and improve the efficiency of the RF system. It will also result in substantial size, weight, and cost reductions in power and cooling systems compared to current solid-state solutions. The rest of the paper focuses on the conceptual design of a 10 MeV, 1 megawatt (MW) beam power module with a 4 ½ cell, 650-MHz, elliptical Nb cavity, coated inside with $Nb_3Sn$ shown in Fig 1. It comprises three main sections: injector, cryostat (main accelerating section), and beam delivery system.

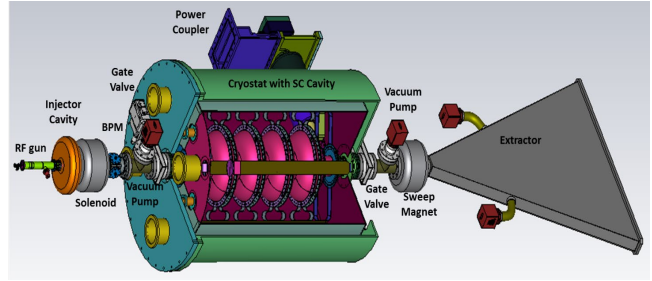

**Figure 1:** Schematic layout of a 1 MW, 10 MeV linac module. Ten identical modules could be used to deliver a 10 MW beam.

## 2. INJECTOR

The proposed external injection scheme includes a thermionic cathode-based RF gun and an injector cavity that provides a pre-acceleration to the beam before entering the superconducting section. The RF gun resonator is installed at the base of the cathode-grid assembly. Note that a grid is used to gate the electron emission. A single cell injector cavity is placed just downstream of the gun-grid assembly. The injector section must deliver a high-quality beam to the SRF section. Thus, the gun design facilitates an operation at a fundamental frequency of 650 MHz and its second harmonic if needed to obtain a short bunch length. To match injection energy at the main accelerating section entrance, a special RF gun and injection cavity design are made to accelerate non-relativistic low energy electrons. In addition, operating parameters such as DC voltage, RF voltage, phase interval of the gun were also optimized to deliver a well-focused beam.

### 2.1. RF Gun

Operational experience in Stanford, Boeing, and Novosibirsk FEL injectors FELs [5] showed that RF guns, based on a thermionic cathode, are cost-effective, simple, and reliable. They can be operated in various regimes with a minimal backward bombardment of electrons and deliver a high-quality beam with low energy and phase spread among particles. In addition, they are ideal for applications that require a high average beam current and long lifetime. These features make the thermionic-cathode RF gun a preferred choice for MW applications. The design has three independent parts: RF resonator with power coupler, thermionic cathode, and grid assembly. The proposed design concept facilitates easy maintenance and assembly. In addition, it allows a rapid repairing/replacement of a particular part during operation and, therefore, improves beam availability during operational hours. Beam availability and reliability are vital features of a successful accelerator facility. The standard series barium tungsten dispenser cathode with a diameter of 0.5 inch at operation temperature 950-1200 $^0$C is commercially available.

### 2.2. Electromagnetic Design of Gun resonator

RF design of the Gun resonator was performed using COMSOL, and results are shown in Fig 2. It operates at a frequency of 650 MHz and a nominal voltage of 2.85 kV. Note that the magnetic field is mainly concentrated in the inner conductor while the peak electric field was obtained in the proximity of the power coupler tip. The plunger position in the gun cavity is adjustable and can be used for fine-tuning the cavity resonance frequency.

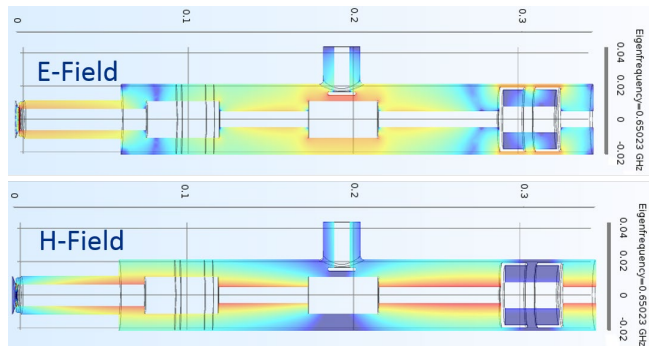

**Figure 2: Electric (top) and magnetic (bottom) fields map in the RF Gun resonator.**

### 2.3. Injector Cavity

An injector cavity is deployed to provide a pre-acceleration to the beam. This is required to have a sufficient beam velocity at the entrance of a β~1 SRF cavity and maximize the beam capture through it. Thus, the primary objective of the optimization was to obtain a maximal accelerating gradient with manageable power loss. Furthermore, the design also accounted for feed-backs from the beam dynamic simulation to determine an optimal accelerating gradient needed from the injection cavity.

## 3. MAIN ACCELERATING SECTION:

The main accelerating section consists of elliptical cells that are conduction-cooled. We show an example of conduction-cooled cell geometry in Fig 3 (right), and the references in this paper go into detail about conduction cooling. Figure 3 also shows the 3-σ beam envelopes in transverse and longitudinal planes. It could be observed from the figure that the beam coming out from the RF-gun section is diverging very rapidly in transverse planes. Thus, a solenoid was placed immediately after the injection section. It provided initial transverse focusing and matched the beam at the cavity entrance. The cavity was set to a negative RF phase relative to crest so that earlier particles experience a lower accelerating field than later particles. It leads to a velocity modulation and, therefore, bunching of the beam. The beam became almost relativistic immediately after the first cell of the cavity. Thus, its longitudinal size remained constant for the rest of the section. The accelerating field in the first cavity cell was higher than for the rest of the cells. This is done to avoid phase slippage and maximize the beam capture across the cavity. The simulation indicates final energy of 10 MeV.

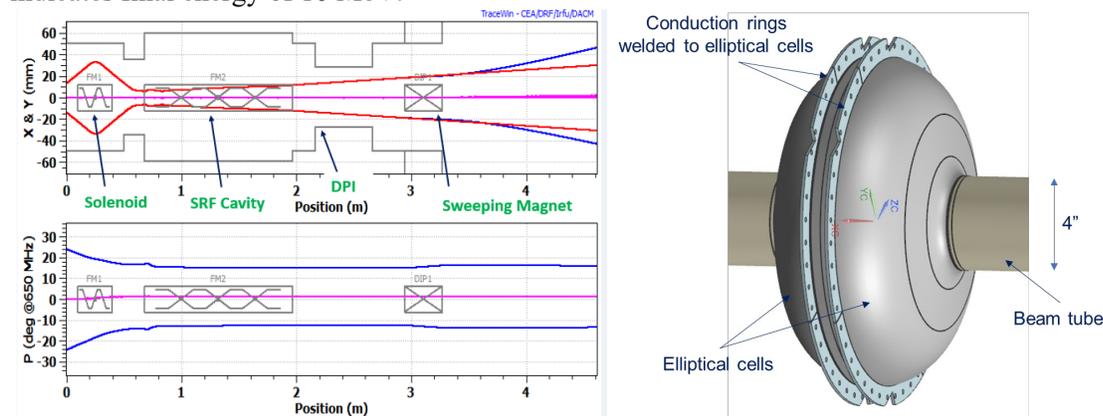

**Figure 3: (a) 3-σ beam envelopes in (top) horizontal (blue), vertical (red), and (bottom) longitudinal planes along the beam transport line. (b) conduction rings welded to the cavity**

## 3. SHIELDING & BEAM DELIVERY SYSTEM

A shielding study was performed using MARSH16 to design an extraction system for 100 mA, 10 MeV linac. The downstream window was optimized for a minimal beam-induced energy deposition density and back-scattering of particles to the SRF cryomodule. Note that a total heat load contribution from the back-scattered particles should not exceed more than 1 W at 4K and 10 W at 70 K to continue a reliable operation of the linac. In addition, cooling, mechanical, and longevity aspects were also considered in this optimization during our complete design study [6].

## 4. CONCLUSIONS

In summary, we have designed a low loss, 10-MeV, 10-MW, industrial electron accelerator that consists of ten modules, each containing a 4 ½ cell, 650-MHz, elliptical Nb cavity, coated inside with $Nb_3Sn$. Each independent module will deliver 1 MW of 10 MeV electrons in a Continuous Wave (CW) operation. This modular approach is attractive to industrial applications because it can provide a straightforward path to scaling to higher powers and provide operational reliability via redundant modules.